\documentstyle[a4,11pt]{article}
\begin{document}
\begin{titlepage}
\vskip 2cm
\begin{flushright}
Preprint CNLP-1997-03
\end{flushright}
\vskip 2cm
\begin{center}
{\bf
SINGULARITY STRUCTURE ANALYSIS, INTEGRABILITY, SOLITONS and DROMIONS in
(2+1)-DIMENSIONAL ZAKHAROV EQUATIONS}\footnote{Preprint
CNLP-1997-03.Alma-Ata.1997 \\
E-mail: cnlpmyra@satsun.sci.kz}
\end{center}
\vskip 2cm
\begin{center}
{\bf  Ratbay MYRZAKULOV}
\end{center}
\vskip 1cm

\begin{center}
 Centre for Nonlinear Problems, PO Box 30, 480035, Alma-Ata-35, Kazakhstan
\end{center}

\begin{abstract}
In this paper, a singularity structure analysis of the (2+1)-dimensional Zakharov
equation is carried out and it is shown that it admits
the Painleve property. The bilinear form of this equation is derived
from the Painleve analysis in a straightforward manner. Using this
bilinear form, we have constructed the simply one soliton solution by the
Hirota method. We have then presented the localized solution (dromion).
The  (2+1)-dimensional Fokas equation is shown to be nothing but the
particular case of the Zakharov equation.
Finally, we have presented the associated integrable spin equations.
\end{abstract}


\end{titlepage}

\setcounter{page}{1}
\newpage

\tableofcontents

\section{Introduction}

In recent years there has been considerable interest in (2+1)-dimensional
soliton equations [1-2]. The study of these equations has thrown up new
ideas in soliton theory, because, they, though, are much richer than their
1+1 dimensional counterparts.  Particularly, the introduction of
exponentially  localized structures (dromions) has triggered renewed interest
in these integrable equations [3-12]. An specially important subject related to
the study of nonlinear differential equations (NLDE) is that concerning the
singularity structure of them [13-14]. The singularity structure analysis appears
as a  systematic procedure for constructing B${\ddot a}$cklund, Darboux
and Miura transformations, Lax representations, different types solutions
etc of given NLDE [14]. At the same time, the Painleve test allows us identify
integrable equations (see, e.g., [3] and refs therein). Notable amongst (2+1)-dimensional soliton equations
are the Davey-Stewartson (DS) equation, the Zakharov equations (ZE), the
Nizhnik-Novikov-Veselov equation, the Ishimori equation, the
Kadomtsev-Petviashvili equation and so on.

In this paper, we consider the following  ZE
$$
iq_{t}+M_{1}q+vq=0 \eqno(1a)
$$
$$
ip_{t}-M_{1}p-vp=0  \eqno(1b)
$$
$$
M_{2} v = -2 M_{1} (pq) \eqno (1c)
$$
where
$$
M_1= \alpha ^2\frac{\partial ^2}{\partial y^2}+4\alpha (b-a)\frac{\partial^2}
   {\partial x \partial y}+4(a^2-2ab-b)\frac{\partial^2}{\partial x^2},
$$
$$
M_2=\alpha^2\frac{\partial^2}{\partial y^2} -2\alpha(2a+1)\frac{\partial^2}
   {\partial x \partial y}+4a(a+1)\frac{\partial^2}{\partial x^2}.
$$
This equation was may  be introduced in [5] and is integrable. Equation
(1) contains several important particular cases. So, for example,
we have the following cases:

(i) $a=b=-\frac{1}{2},$ it yields the DS equation
$$
iq_t + q_{xx} + \alpha^{2}q_{yy} + vq = 0 \eqno(2a)
$$
$$
\alpha^{2}v_{yy} - v_{xx}=-2(\alpha^{2}(pq)_{yy} + (pq)_{xx}). \eqno(2b)
$$

(ii) When $a=b=-1$, we obtain the following equation
$$
iq_{t}+ q_{YY} + vq = 0 \eqno(3a)
$$
$$
ip_{t}-p_{YY} - vp = 0 \eqno(3b)
$$
$$
v_{X} + v_{Y} + 2(pq)_{Y} =0  \eqno(3c)
$$
where $X=x/2, \quad Y = y/\alpha $.

(iii) If $a=b=-1, X=t$, then equation (1) reduces to the following
(1+1)-dimensional Yajima-Oikawa equation [19]
$$
iq_{t}+ q_{YY} + vq = 0 \eqno(4a)
$$
$$
ip_{t}-p_{YY} - vp = 0  \eqno(4b)
$$
$$
v_{t} + v_{Y} + 2(pq)_{Y} =0  \eqno(4c)
$$
and so on [16].

Even though the ZE (1) is known to be completely integrable,
its Painleve property has
not yet been established.  Also the interesting question arises naturally,
whether there exist dromion solutions in equation (1) as well.
In this paper, following Lakshmanan and coworkers (see, e.g. [3] and
references therein),
we address ourselves to these problems and carry out
the singularity structure analysis  and confirm its Painleve nature.
We also deduce its bilinear form from the Painleve analysis. Next we
construct soliton and dromion solutions using the Hirota method.
We also show that the Fokas equation (FE) is the particular case of the
ZE as $a=-\frac{1}{2}, \alpha^{2} = 1.$

The present work falls into seven parts. In section II, we present the
equivalent forms  of the ZE (1) apart from studing its properties. In
section III, we carry out the singularity structure analysis of equation (1)
and confirm its Painleve nature. We then obtain the Hirota bilinear form
directly from the Painleve analysis in section IV.  In section V, we
generate the simplest one soliton solution (1-SS),  the dromion  and
1-rational solutions.
A connection between the ZE and the FE we will discuss
in section VI. The associated integrable spin systems we present in section VII.
Section VIII contains a short discussion
of the results.

\section{Lax representation and equivalent forms}

Equation (1) has the following Lax representation [5]
$$
\alpha \Psi_y =2B_{1}\Psi_x + B_{0}\Psi \eqno(5a)
$$
$$
\Psi_t=4iC_{2}\Psi_{xx}+2C_{1}\Psi_x+C_{0}\Psi \eqno(5b)
$$
with
$$
B_{1}= \pmatrix{
a+1 & 0 \cr
0   & a
},\quad
B_{0}= \pmatrix{
0   &  q \cr
p   &  0
}
$$
$$
C_{2}= \pmatrix{
b+1 & 0 \cr
0   & b
},\quad
C_{1}= \pmatrix{
0   &  iq \cr
ip  &  0
},\quad
C_{0}= \pmatrix{
c_{11}  &  c_{12} \cr
c_{21}  &  c_{22}
}
$$
$$
c_{12}=i[2(2b-a+1)q_{x}+i\alpha q_{y}] \quad
c_{21}=i[2(a-2b)p_{x}-i\alpha p_{y}]
$$
and $v=i(c_{22}-c_{11})$. Here $c_{jj}$ are the solution of the  following
equations
$$
2(a+1) c_{11x}- \alpha c_{11y} = i[2(2b-a+1)(pq)_{x} + \alpha (pq)_{y}]
$$
$$
2ac_{22x}-\alpha c_{22y} = i[2(a-2b)(pq)_{x} - \alpha (pq)_{y}].
$$

For our future algebra the ZE in the form (1) is rather complicated.
Because it has sense to look for a other more convenient and elegant forms
of equation (1). The first form we obtain from the compatibility condition
of  equations (5). We have
$$
iq_{t}+M_{1}q+i(c_{22}-c_{11})q=0 \eqno(6a)
$$
$$
ip_{t}-M_{1}p-i(c_{11}-c_{22})p=0  \eqno(6b)
$$
$$
2(a+1) c_{11x}- \alpha c_{11y} = i[2(2b-a+1)(pq)_{x} + \alpha (pq)_{y}]
\eqno (6c)
$$
$$
2ac_{22x}-\alpha c_{22y} = i[2(a-2b)(pq)_{x} - \alpha (pq)_{y}].
\eqno (6d)
$$
Now, if we introduce the following transformations
$$
V^{\prime} = c_{22} - i(2b+1)pq ,  \quad U^{\prime} = c_{11} -i(2b+1) pq
\eqno (7)
$$
then equation (6) takes the form
$$
iq_{t}+M_{1}q+i(V^{\prime} - U^{\prime}) q = 0 \eqno(8a)
$$
$$
ip_{t}-M_{1}p-i(V^{\prime} - U^{\prime}) p = 0 \eqno(8b)
$$
$$
2a V^{\prime}_{x}- \alpha V^{\prime}_{y} = -2ib[2(a+1)(pq)_{x} -
\alpha (pq)_{y}] \eqno (8c)
$$
$$
2(a+1)U^{\prime}_{x}-\alpha U^{\prime}_{y} = -2i(b+1)[2a(pq)_{x} -
\alpha (pq)_{y}]. \eqno (8d)
$$
Let us we now  rewrite this equation in the following form
$$
iq_t+ (1 + b)q_{\xi \xi } - b q_{\eta \eta } + [2bV-2(b+1)U]q = 0 \eqno(9a)
$$
$$
ip_t - (1 + b)p_{\xi \xi } + b q_{\eta \eta } - [2bV-2(b+1)U]p = 0 \eqno(9b)
$$
$$
V_{\xi}  = (pq)_{\eta} \eqno (9c)
$$
$$
U_{\eta} = (pq)_{\xi} \eqno (9d)
$$
where $U, V, \xi$ and  $\eta $ are defined by
$$
U^{\prime} = -2i(b+1)U, \quad V^{\prime} = -2ibV, \quad
\xi = \frac{x}{2} + \frac{a+1}{\alpha}y, \quad \eta = -\frac{x}{2} -
\frac{a}{\alpha}y.    \eqno(10)
$$
Having this form of the ZE, we are in a convenient position to explore the
singularity structure of it. Note that in terms of $\xi, \eta$, equation (1)
takes the form
$$
iq_t+ (1 + b)q_{\xi \xi } - b q_{\eta \eta } + vq = 0 \eqno(11a)
$$
$$
ip_t - (1 + b)p_{\xi \xi } + b q_{\eta \eta } -  vq = 0 \eqno(11b)
$$
$$
v_{\xi \eta } = -2[(1+ b) (pq)_{\xi \xi} - b(pq)_{\eta \eta}]. \eqno(11c)
$$
In particular, from this equation as $b=0$, we get the other ZE [5]
$$
iq_t+ q_{\xi \xi } + vq = 0 \eqno(12a)
$$
$$
ip_t - p_{\xi \xi } -  vq = 0 \eqno(12b)
$$
$$
v_{\eta } = -2 (pq)_{\xi }. \eqno(12c)
$$

\section{Singularity structure analysis}

In order to carry out a singularity structure analysis, following [3],
we effect
a local Laurent expansion
in the neighbourhood
of a noncharacteristic singular manifold $\phi(\xi, \eta, t) = 0,
\quad (\phi_{\xi}, \phi_{\eta}, \phi_{t} \ne 0)$. We assume the  leading
orders of the solutions of equation (9) to take  the form
$$
q=q_{0}\phi^{m}, \quad p=p_{0}\phi^{n}, \quad V=V_{0}\phi^{\gamma},
\quad U=U_{0}\phi^{\delta}  \eqno(13)
$$
where $q_{0}, \quad p_{0}, \quad V_{0}$ and $U_{0}$ are analytic functions
of $(\xi, \eta, t)$. In (13)  $ m, n, \gamma$ and $\delta$ are integers
(if they exist) to be evaluated.  Substituting expressions (13) into equation
(9) and balancing the most dominant terms, we get
$$
m = n = - 1, \quad \gamma = \delta = - 2 \eqno(14)
$$
and  the following equations
$$
p_{0}q_{0} = \phi_{\xi}\phi_{\eta}, \quad V_{0} = \phi^{2}_{\eta},
\quad U_{0} = \phi^{2}_{\xi}.
\eqno(15)
$$
To evaluate the resonances, we consider the Laurent series of the
solutions
$$
q=\sum_{j=0}q_{j}\phi^{j-1}, \quad
p=\sum_{j=0}p_{j}\phi^{j-1}, \quad
V=\sum_{j=0}V_{j}\phi^{j-2}, \quad
U=\sum_{j=0}U_{j}\phi^{j-2}. \eqno(16)
$$
Then we substitute these expansions into equation (9) and equate the
coefficients $(\phi^{j-3}, \phi^{j-3}, \phi^{j-3},$\\$ \phi^{j-3})$ to zero
to give
$$
\left ( \begin{array}{cccc}
j(j-3)[(b+1)\phi_{\xi}^{2}-b\phi_{\eta}^{2}]  &  0  & 2bq_{0} &-2(b+1)q_{0} \\
0 & j(j-3)[(b+1)\phi_{\xi}^{2}-b\phi_{\eta}^{2}]  &  2bp_{0}  & -2(b+1)p_{0}  \\
(j-2)p_{0}\phi_{\eta}   &  (j-2)q_{0}\phi_{\eta}  & - (j- 2)\phi_{\xi} & 0  \\
(j-2)p_{0}\phi_{\xi}    & (j-2)q_{0}\phi_{\xi}    &  0   & - (j-2)\phi_{\eta}
\end{array} \right )
\left ( \begin{array}{c}
q_{j} \\
p_{j} \\
V_{j} \\
U_{j}
\end{array} \right )
= 0 \eqno(17)
$$
From the condition for the existence of nontrivial solutions to equation (17),
we get the resonance values as
$$
j = - 1, 0, 2, 2, 4.    \eqno(18)
$$

Obviously, the resonance at $j = - 1$ represents the arbitrariness
of the singularity manifold $\phi(\xi, \eta, t) = 0$. At the same time,
the resonance at $j=0$ is associated with the arbitrariness of the
functions  $q_{0}, p_{0}, V_{0}$ or $U_{0}$ (cf equation (15)).
To prove the existence of arbitrary functions at the other resonance values
$j=2,2,3,4$, we use the Laurent expansion (16) into equation (9).

Now, gathering  the coefficients of  $(\phi^{-2}, \phi^{-2}, \phi^{-2},
\phi^{-2})$, we obtain
$$
2[bV_{0} -(b+1)U_{0}]q_{1} +2bq_{0}V_{1} - 2(b+1)q_{0}U_{1} = A_{1} \eqno(19a)
$$
$$
2[bV_{0} -(b+1)U_{0}]p_{1} +2bp_{0}V_{1} - 2(b+1)p_{0}U_{1} = B_{1} \eqno(19b)
$$
$$
p_{0} \phi_{\eta} q_{1} + q_{0} \phi_{\eta} p_{1} - \phi_{\xi} V_{1}
= C_{1}   \eqno(19c)
$$
$$
p_{0} \phi_{\xi} q_{1} + q_{0} \phi_{\xi} p_{1} - \phi_{\eta} U_{1}
= D_{1}   \eqno(19d)
$$
where
$$
A_{1} = iq_{0} \phi_{t} + (b+1)[2q_{0 \xi} \phi_{\xi} +
q_{0} \phi_{\xi \xi}] -b[2q_{0 \eta} \phi_{\eta} +
q_{0} \phi_{\eta \eta}]        \eqno(20a)
$$
$$
B_{1} = -ip_{0} \phi_{t} + (b+1)[2p_{0 \xi} \phi_{\xi} +
p_{0} \phi_{\xi \xi}] -b[2p_{0 \eta} \phi_{\eta} +
p_{0} \phi_{\eta \eta}]        \eqno(20b)
$$
$$
C_{1} =\phi_{\xi} \phi_{\eta \eta} - \phi_{\eta} \phi_{\xi \eta} \eqno (20c)
$$
$$
D_{1} = \phi_{\eta} \phi_{\xi \xi} - \phi_{\xi} \phi_{\xi \eta}.  \eqno(20d)
$$
The solution of equation (19) has the form
$$
q_{1} = \frac{iq_{0} \phi_{t} +(b+1)[2q_{0 \xi} \phi_{\xi} -
q_{0} \phi_{\xi \xi}]-b[2q_{0\eta} \phi_{\eta}
-q_{0} \phi_{\eta \eta}]}{2[b \phi_{\eta}^{2} - (b+1) \phi_{\xi}^{2}]}
\eqno (21a)
$$
$$
p_{1} = \frac{-ip_{0} \phi_{t} +(b+1)[2p_{0 \xi} \phi_{\xi} -
p_{0} \phi_{\xi \xi}]-b[2p_{0\eta} \phi_{\eta}
-p_{0} \phi_{\eta \eta}]}{2[b \phi_{\eta}^{2} - (b+1) \phi_{\xi}^{2}]}
\eqno (21b)
$$
$$
V_{1} = - \phi_{\eta \eta}   \eqno   (21c)
$$
$$
U_{1} = - \phi_{\xi \xi}.    \eqno  (21d)
$$
Similarly, collecting the coefficients of $ (\phi^{-1}, \phi^{-1}, \phi^{-1},
\phi^{-1} ),$ we obtain
$$
2[bV_{0} - (b+1)U_{0}]q_{2} + 2bq_{0}V_{2} -2(b+1)q_{0}U_{2} = A_{2}
\eqno  (22a)
$$
$$
2[bV_{0} - (b+1)U_{0}]p_{2} + 2bp_{0}V_{2} -2(b+1)p_{0}U_{2} = B_{2}
\eqno  (22b)
$$
$$
V_{1 \xi} = (p_{0} q_{1} + p_{1} q_{0})_{\eta}   \eqno   (22c)
$$
$$
U_{1 \eta} = (p_{0} q_{1} + p_{1} q_{0})_{\xi}   \eqno   (22d)
$$
where
$$
A_{2} = -iq_{0t} +bq_{0 \eta \eta} -
(b+1)q_{0 \xi \xi} -[2bV_{1} -2(b+1)U_{1}]q_{1}     \eqno   (23a)
$$
$$
B_{2} = ip_{0t} +bp_{0 \eta \eta} -
(b+1)p_{0 \xi \xi} -[2bV_{1} -2(b+1)U_{1}]p_{1}.     \eqno   (23b)
$$
Equations (22c,d) are identically satisfied. So, we have
only two equations for four unknown functions $a_{2}, b_{2},
V_{2}, U_{2} $, i.e.,  two of them must be arbitrary.

Now, collecting  the coefficients of $ (\phi^{0}, \phi^{0}, \phi^{0},
\phi^{0}),$ we have
$$
2[bV_{3} - (b+1)U_{3} ]q_{0} = A_{3}    \eqno   (24a)
$$
$$
2[bV_{3} - (b+1)U_{3} ]p_{0} = B_{3}    \eqno   (24b)
$$
$$
\phi_{\eta}(p_{3}q_{0}  + p_{0} q_{3}) - V_{3} \phi_{\xi} = C_{3}  \eqno  (24c)
$$
$$
\phi_{\xi}(p_{3}q_{0}  + p_{0} q_{3}) - U_{3} \phi_{\eta} = D_{3}  \eqno  (24d)
$$
with
$$
A_{3} = -iq_{1t} - iq_{2} \phi_{t} -(b+1)[q_{1 \xi \xi} + 2q_{2 \xi} \phi_{\xi} +
q_{2} \phi_{\xi \xi}]+b[q_{1 \eta \eta} + 2q_{2 \eta} \phi_{\eta} +
q_{2} \phi_{\eta \eta}]
$$
$$
-2b[V_{1}q_{2} +V_{2}q_{1}] +2(b+1)[U_{1}q_{2} + U_{2}q_{1}]  \eqno  (25a)
$$
$$
B_{3} = ip_{1t} + ip_{2} \phi_{t} -(b+1)[p_{1 \xi \xi} + 2p_{2 \xi} \phi_{\xi} +
p_{2} \phi_{\xi \xi}]+b[p_{1 \eta \eta} + 2p_{2 \eta} \phi_{\eta} +
p_{2} \phi_{\eta \eta}]
$$
$$
-2b[V_{1}p_{2} +V_{2}p_{1}] +2(b+1)[U_{1}p_{2} + U_{2}p_{1}]  \eqno  (25b)
$$
$$
C_{3}= V_{2\xi} - [(p_{0}q_{2})_{\eta}+(p_{1}q_{1})_{\eta}+(p_{2}q_{0})_{\eta} +
(p_{2}q_{1} + p_{1}q_{2}) \phi_{\eta}]     \eqno  (25c)
$$
$$
D_{3}= U_{2\eta} - [(p_{0}q_{2})_{\xi}+(p_{1}q_{1})_{\xi}+(p_{2}q_{0})_{\xi} +
(p_{2}q_{1} + p_{1}q_{2}) \phi_{\xi}].     \eqno  (25d)
$$

This system can be reduced to the three equations in four unknown functions,
hence follows that  one of the functions $q_{3}, p_{3}, V_{3} $ and $U_{3}$
is arbitrary. Proceeding further
to the coefficients of $(\phi^{1}, \phi^{1}, \phi^{1}, \phi^{1})$ ,
we have checked that one of the functions $q_{4}, p_{4}, V_{4}$ and $U_{4}$
is arbitrary. Thus the general solution $(q, p, V, U) (\xi, \eta, t)$ of
equation (9) admits  the required number of arbitrary functions without
the introduction of any movable critical manifold, thereby passing
the Painleve property. Thus the ZE (9) is expected to be integrable.

\section {Bilinearization and B${\ddot a}$cklund transformation}

Using the results of the previous section,  we can investigate the other
integrability properties of equation (9). Particularly,
we can  construct B${\ddot a}$cklund, Darboux
and Miura transformations, Lax representations, bilinear form,
different types solutions of the ZE. For example, to obtain
the B${\ddot a}$cklund transformation  of equation (9),
we truncate the Laurent series at the constant level term, that is
$$
q_{j-1} = p_{j-1} = V_{j} = U_{j} = 0, \quad j \geq 3  \eqno(26)
$$
In this case, from (16) we have
$$
q = q_{0}\phi^{-1} + q_{1}, \quad  p = p_{0}\phi^{-1} + p_{1}  \eqno(27a)
$$
$$
V=V_{0}\phi^{-2} +V_{1}\phi^{-1} + V_{2}, \quad U=U_{0}\phi^{-2}+
U_{1}\phi^{-1}+U_{2} \eqno(27b)
$$
where $(q, q_{1}), (p, p_{1}), (V, V_{2})$ and $(U, U_{2})$ satisfy equation
(9) with $(q_{0}, p_{0}, V_{0}, U_{0})$ and $(V_{1}, U_{1})$ satisfying
equations (15). If we take the vacuum
solution $q_{1}=p_{1}=V_{2}=U_{2}=0 $, then from  the above
B${\ddot a}$cklund transformation (27) we have
$$
q = q_{0}\phi^{-1}  \eqno   (28a)
$$
$$
p = p_{0}\phi^{-1}   \eqno   (28b)
$$
$$
V = V_{0}\phi^{-2} + V_{1}\phi^{-1} = - \partial_{\eta \eta}\log\phi
\eqno (28c)
$$
$$
U = U_{0}\phi^{-2} + U_{1}\phi^{-1} = - \partial_{\xi \xi}\log\phi.
\eqno (28d)
$$
Hence and from (9), in the case, when $\phi$ is real,  follows
$$
[iD_{t} + (b+1)D_{\xi}^{2} - bD_{\eta}^{2}]q_{0}\circ \phi = 0   \eqno  (29a)
$$
$$
[iD_{t} - (b+1)D_{\xi}^{2} + bD_{\eta}^{2}]p_{0}\circ \phi = 0   \eqno  (29b)
$$
$$
D_{\xi}D_{\eta} \phi \circ \phi = -2p_{0}q_{0}       \eqno  (29c)
$$
which is the desired Hirota bilinear form for equations (9). Note that the
bilinear form of the ZE for its form (1) is given by
$$
[iD_{t} - 4(a^{2} -2ab -b)D_{x}^{2} - 4 \alpha (b-a)D_{x} D_{y} -
\alpha^{2} D_{y}^{2} ](G \circ \phi) = 0  \eqno (29e)
$$
$$
[iD_{t} - 4(a^{2} -2ab -b)D_{x}^{2} - 4 \alpha (b-a)D_{x} D_{y} -
\alpha^{2} D_{y}^{2} ](P\circ \phi) = 0  \eqno (29f)
$$
$$
[4a(a+1)D_{x}^{2} - 2\alpha (2a+1)D_{x} D_{y} + \alpha^{2} D_{y}^{2})
(\phi \circ \phi) = -2PG  \eqno (29g)
$$
where
$$
q = \frac{G}{\phi}, \quad p = \frac{P}{\phi} \eqno(29h)
$$
with
$$
v=2M_{2}\log \phi .   \eqno(29i)
$$
Hereafter $(29)\equiv (29a,b,c).$

\section{Simplest solutions}

Equations (29) allow us to obtain the interesting classes of
solutions for the ZE (9) [25]. Below we find some simplest solutions
of equation (9), when  $p=Eq^{*}, E=\pm 1, \alpha^{2}=1$.
In this case, the Hirota bilinear
equations (29) take the form $(q_{0} \equiv g)$
$$
[iD_{t} + (b+1)D_{\xi}^{2} - bD_{\eta}^{2}]g\circ \phi = 0   \eqno  (30a)
$$
$$
D_{\xi}D_{\eta} \phi \circ \phi = -2Egg^{*}       \eqno  (30b)
$$

\subsection{The 1-soliton  solution}

The construction of the  soliton solutions is standard.
One expands the functions $g$ and $\phi$ as a series of $\epsilon$
$$
g = \epsilon g_{1} + \epsilon^{3} g_{3} + \epsilon^{5}g_{5} +
\cdot \cdot \cdot \cdot \cdot \eqno(31a)
$$
$$
\phi =1+\epsilon^2 \phi_2+\epsilon^4 \phi_4+\epsilon^6 \phi_6 + .....\quad .
\eqno(31b)
$$

Substituting these expansions into (30) and equating the coefficients
of $\epsilon $, in the 1-SS case, one obtains the following system
of equations:
$$
\epsilon^1: \quad [iD_{t} + (b+1)D_{\xi}^{2} - bD_{\eta}^{2}]g_{1} \circ 1 = 0
\eqno  (32a)
$$
$$
\epsilon^3: \quad [iD_{t} + (b+1)D_{\xi}^{2} - bD_{\eta}^{2}]g_{1} \circ \phi_{2}
= 0   \eqno  (32b)
$$
$$
\epsilon^{2}: \quad D_{\xi}D_{\eta} (1\circ \phi_{2} + \phi_{2} \circ 1) = -2Egg^{*}  \eqno  (32c)
$$
$$
\epsilon^{4}: \quad D_{\xi}D_{\eta} (\phi_{2} \circ  \phi_{2}) = 0.  \eqno  (32d)
$$
Using these equations  we can  construct the 1-SS
of equation (5). In order to construct exact 1-SS of equation (9),
we take the ansatz
$$
g_1= \exp {\chi_1} \eqno (33)
$$
where
$$
\chi_1 = p_1\xi + s_1\eta + c_1t + e_1, \quad p_{1} = p_{1R} + ip_{1I},
\quad  s_{1}=s_{1R}+is_{1I}. \eqno(34)
$$
Sustituting (33) into (32a), we obtain
$$
c_1=i[(b+1)p^{2}_{1} - bs^{2}_{1}].   \eqno  (35)
$$

The expression for $\phi_2$ , we get from (32c)
$$
\phi_2=\exp{(\chi_1+\chi_1^* + 2\psi)} \eqno(36)
$$
with
$$
\exp(2\psi) = -E/4p_{1R}s_{1R}.          \eqno(37)
$$
Equations (32b,d) are identically satisfied. Finally, from (28), (33) and (36),
we get the 1-SS of equation (9)
in the form
$$
q(\xi, \eta, t) = \frac{1}{2}\exp(-\psi)sech(\chi_{1R} +
\psi)\exp(i\chi_{1I})       \eqno  (38a)
$$
$$
V(\xi, \eta, t) = -s_{1R}^{2}sech^{2}(\chi_{1R} +\psi)      \eqno (38b)
$$
$$
U(\xi, \eta, t) = -p_{1R}^{2}sech^{2}(\chi_{1R} +\psi)      \eqno   (38c)
$$
and for the hybrid potential
$$
v(\xi, \eta, t) = 2[(b+1)p^{2}_{1R}-bs_{1R}^{2}]sech^{2}(\chi_{1R} +\psi)
\eqno (38d)
$$
where $\chi_{1R} = Re \chi_{1} = p_{1R}\xi + s_{1R}\eta -
[2(b+1)p_{1R}p_{1I}-2bs_{1R}s_{1I}]t.$  This algebra  can be
used to construct $N-$line soliton solutions as well. As shown in [3],
the above 1-SS reveals the fact that
$$
q\rightarrow 0, \quad U\rightarrow 0, \quad
V = -s_{1R}^{2}sech^{2}(\chi^{\prime}_{1R} +\psi{\prime})\rightarrow
v_{1}(\eta, t) \quad as \quad p_{1R} \rightarrow 0 \eqno (39)
$$
where $\chi^{\prime}_{1R} = s_{1R}[\eta + 2bs_{1I}t] + e_{1R}$ and $
\psi^{\prime}$ is a new phase constant. Similarly, we have
$$
q\rightarrow 0, \quad V\rightarrow 0, \quad
U = -p_{1R}^{2}sech^{2}(\chi^{\prime \prime}_{1R} +
\psi{\prime \prime})\rightarrow
u_{1}(\eta, t) \quad as \quad p_{1R} \rightarrow 0 \eqno (40)
$$
where $
\chi^{\prime \prime}_{1R} = p_{1R}[\xi - 2(b+1)p_{1I}t] + e_{1R}$ and $
\psi^{\prime\prime}$ is another phase constant.
Thus, as in [3],   the solution is composed of
two ghost solitons $v_{1}(\eta, t) $ and $u_{1}(\xi, t)$ driving the
potentials $V$ and $U$ respectively in the absence of the physical field $q$.
Note that these results are the same as in [3].

\subsection{The (1,1)-dromion   solutions}

Let us we now construct a dromion solution of the ZE. For example,
to get a simple (1, 1) dromion solution, following Radha and Lakshmanan (see,
e.g [3]), we take the ansatz
$$
g_{11D}=\rho\exp(\chi_{1} + \chi_{2})   \eqno (41a)
$$
$$
\phi_{11D} = 1+ j\exp(\chi_{1} + \chi_{1}^{*}) + k\exp(\chi_{2} + \chi_{2}^{*})
 + l\exp(\chi_{1} + \chi_{1}^{*} +\chi_{2} + \chi_{2}^{*})    \eqno  (41b)
$$
where $j, k, l$ are real positive constants and
$$
\chi_{1} = p_{1} \xi + i(b+1)p_{1}^{2}t + \chi_{1}^{0}, \quad
\chi_{2} = s_{2} \eta - ibs_{2}^{2}t + \chi_{2}^{0}.     \eqno   (42)
$$
Here $p_{1}=p_{1R} + ip_{1I}, s_{2} =s_{2R}+is_{1I}$ are complex constants.
Substituting (41) into (30),  we get the following conditions
$$
\mid \rho \mid ^{2} = 4p_{1R}s_{1R}(jk-l)/E, \quad
(l-jk)\exp(-2\psi)>0. \eqno (43)
$$
At last, from (41) and (28) , we obtain the (1, 1) dromion solution
in the form
$$
q_{11D} = \frac{g_{11D}}{\phi_{11D}}, \quad V_{11D} = -
\partial_{\eta \eta}\log \phi_{11D},
\quad U_{11D} =-\partial_{\xi\xi} \log \phi_{11D}  \eqno(44a)
$$
or
$$
q_{11D} = \frac{\rho\exp(\chi_{1} + \chi_{2})}{1+j\exp(\chi_{1}+
\chi_{1}^{*}) + k\exp(\chi_{2}+\chi_{2}^{*}) +
l\exp(\chi_{1} + \chi_{1}^{*} + \chi_{2} + \chi_{2}^{*})}
\eqno(44b)
$$
$$
V_{11D} = \frac{-4s_{2R}^{2}\exp(2\chi_{2R})[(k+l\exp(2\chi_{2R})][(1
+j\exp(2\chi_{1R})]}{[1+j\exp(2\chi_{1R}) +k\exp(2\chi_{2R}) +
l\exp(2(\chi_{1R}+\chi_{2R}))]^{2}} \eqno(44c)
$$
$$
U_{11D} = \frac{-4p_{1R}^{2}\exp(2\chi_{1R})[j+l\exp(2\chi_{1R})][1
+k\exp(2\chi_{2R})]}{[1+j\exp(2\chi_{1R}) +k\exp(2\chi_{2R})
 +l\exp(2(\chi_{1R}+\chi_{2R}))]^{2}}.      \eqno(44d)
$$
From the last two equations, we can get the  expression for the hybrid
potential $v$ (9)
$$
v_{11D} = 2bV_{11D} - 2(b+1)U_{11D}.  \eqno(44e)
$$

\subsection{The 1-rational solution}
In this subsection, we want present the simple 1-rational
solution of equation (9). Let $g_{1}=b_{0}=const$. Then, from (32) we get
$$
\phi_{2} = -E|b_{0}|^{2}\xi\eta + b_{1}\eta, \quad b_{1}=const. \eqno(45)
$$
So, the 1-rational solution has the form
$$
q = \frac{b_{0}}{1-E|b_{0}|^{2}\xi\eta + b_{1}\eta} \eqno (46a)
$$
$$
V = [\frac{b_{1}-E|b_{0}|^{2}\xi}{1-E|b_{0}|^{2}\xi\eta +
 b_{1}\eta}]^{2} \eqno (46b)
$$
$$
U = [\frac{|b_{0}|^{4}\eta^{2}}{1-E|b_{0}|^{2}\xi\eta +
 b_{1}\eta}]^{2}. \eqno (46c)
$$
Note that in this case, we have the following boundary conditions
$$
(q, U, V)|_{\xi = \pm \infty}  =  (0,0, \frac{1}{\eta^{2}}=v_{2}(\eta))
 \eqno (47a)
$$
and
$$
(q, U, V)|_{\eta = \pm \infty}  =  (0, \frac{1}{\xi^{2}}=u_{2}(\xi), 0).
 \eqno (47b)
$$

\section{A connection between the ZE and the FE}

Now let us we consider the FE [4]
$$
iq_t - (\gamma - \beta)q_{\xi \xi } +  (\gamma + \beta)
q_{\eta \eta } - 2\lambda q[(\gamma +
\beta)(\int^{\xi}_{-\infty}(pq)_{\eta}d\xi^{\prime}
$$
$$
+ v_{1}(\eta, t))
-  (\gamma - \beta)(\int^{\eta}_{-\infty}(pq)_{\xi} d\eta^{\prime}
+ v_{2}(\xi,t))] =0 \eqno(48a)
$$
$$
ip_t + (\gamma - \beta)p_{\xi \xi } -  (\gamma + \beta)
p_{\eta \eta } + 2\lambda p[(\gamma +
\beta)(\int^{\xi}_{-\infty}(pq)_{\eta}d\xi^{\prime}
$$
$$
+ v_{1}(\eta, t))
-  (\gamma - \beta)(\int^{\eta}_{-\infty}(pq)_{\xi} d\eta^{\prime}
+ v_{2}(\xi,t))] =0 \eqno(48b)
$$
with $ p=\bar q$ and  in contrast with the equation (9), in this case
$\xi, \eta$ are the characteristic coordinates defined by
$$
\xi = x+y, \quad \eta = x-y. \eqno(49)
$$

This equation also contains several  interesting particular cases. Let
us recall these cases.

(i) $\gamma = \beta = \frac{1}{2}, v_{1} = v_{2} =0,$ yields equation
$$
iq_t +q_{\eta\eta} - 2\lambda q\int^{\xi}_{-\infty}(pq)_{\eta}d\xi^{\prime}
= 0, \quad \lambda = \pm 1.  \eqno(50)
$$
As noted by Fokas, equation (50) is perhaps the simplest complex
scalar equation in 2+1 dimensions, which can be solved by the IST method.
It is also worth pointing out that when $x=y$ this equation reduces
to the (1+1)-dimensional integrable NLSE.

(ii) $\gamma = 0, \beta = 1$, yields the celebrated DSI equation
$$
iq_t + q_{\xi \xi } +  q_{\eta \eta }
- 2\lambda q[(\int^{\xi}_{-\infty}(pq)_{\eta}d\xi^{\prime} +
v_{1}(\eta, t))
+(\int^{\eta}_{-\infty}(pq)_{\xi} d\eta^{\prime}
+ v_{2}(\xi,t))] =0 .\eqno(51)
$$
This equation has the Painleve property  and admits exponentially
localized solutions including dromions for nonvanishing boundaries.

(iii) $\gamma = 1, \beta =0$ yields the DSIII equation
$$
iq_t - q_{\xi \xi } +  q_{\eta \eta }
- 2\lambda q[(\int^{\xi}_{-\infty}(pq)_{\eta}d\xi^{\prime} +
v_{1}(\eta, t))
-(\int^{\eta}_{-\infty}(pq)_{\xi} d\eta^{\prime}
+ v_{2}(\xi,t))] =0. \eqno(52)
$$
Equation (52)  also supports certain localized solutions.

Now we return to the ZE (9) and make the simplest scaling
tranformation: from $(t,\xi, \eta, q,p,v)$
to $(Ft, C\xi, D\eta, Aq, Bp, HV, EU)$. Then, for example,  equation (9)
takes the form
$$
iq_t - (\gamma - \beta)q_{\xi \xi } +  (\gamma + \beta)
q_{\eta \eta } -2\lambda [(\gamma+\beta)V-(\gamma-\beta)U]q = 0 \eqno(53a)
$$
$$
ip_t + (\gamma - \beta)p_{\xi \xi } -  (\gamma + \beta)
p_{\eta\eta } +2\lambda [(\gamma+\beta)V-(\gamma-\beta)U]p = 0 \eqno(53b)
$$
$$
V_{\xi}  = (pq)_{\eta} \eqno (53c)
$$
$$
U_{\eta} = (pq)_{\xi} \eqno (53d)
$$
where
$$
\lambda =\frac{ABCD}{F},\quad  F= \frac{\beta - \gamma}{1+b}C^{2}, \quad
D^{2}=\frac{b(\gamma-\beta)}{(1+b)(\gamma+\beta)}C^{2},
$$
$$
\gamma = -\frac{1}{2}F[(b+1)D^{2}+bC^{2}]C^{-2}D^{-2}, \quad
\beta = \frac{1}{2}F[(b+1)D^{2}-bC^{2}]C^{-2}D^{-2}.
$$

From (53c,d), we get
$$
V = \int^{\xi}_{-\infty}(pq)_{\eta}d\xi^{\prime} + v_{1} (\eta, t) \eqno (54a)
$$
$$
U = \int^{\eta}_{-\infty}(pq)_{\xi}d\eta^{\prime} + v_{2}(\xi,t). \eqno (54b)
$$
Substituting (54) into (53a,b),  we obtain the FE (48).
Thus, we have proved that the ZE and the FE are  equivalent to
each other, as $\alpha^{2}=1, a=-\frac{1}{2}$.
In particular, this is why the ZE contains and at the same time
the FE not contains the DSII equation.
Recently it was proved by
Radha and Lakshmanan [3] that the FE (48) satisfies the Painleve property and
hence it is expected to be integrable. From these results follow that
the ZE also satisfies the Painleve property and is integrable.

\section{ Associated integrable spin systems}

In this section, we wish present, in a briefly form, the spin equivalent
counterpart of the ZE and its reductions.
It is well known that the ZE (1) is gauge equivalent to the Myrzakulov
IX (M-IX) equation [16]
$$
iS_t + \frac{1}{2}[S, M_1 S] + A_2 S_x + A_1 S_y = 0  \eqno(55a)
$$
$$
M_2u = \frac{\alpha^{2}}{2i}tr( S[ S_x , S_y]) \eqno(55b)
$$
where $ \alpha,b,a  $=  consts,
$$
S= \pmatrix{
S_3 & rS^- \cr
rS^+ & -S_3
},\quad S^{\pm}=S_{1}\pm iS_{2}, \quad  S^2 = EI,\quad E = \pm 1,
\quad r^{2}=\pm 1,
$$
$$
A_1=i\{\alpha (2b+1)u_y - 2(2ab+a+b)u_{x}\},
$$
$$
A_2=i\{4\alpha^{-1}(2a^2b+a^2+2ab+b)u_x - 2(2ab+a+b)u_{y}\}.
$$
This equation is integrable and also admits several integrable reductions.
There some of them:

(i) {\it The Myrzakulov VIII (M-VIII) equation}. First, let us  we consider the
reduction of the M-IX equation (55)
as $ a=b=-1$.
We have [16]
$$
iS_t+\frac{1}{2}[S,S_{YY}]+iwS_Y = 0 \eqno(56a)
$$
$$
w_{X} + w_{Y} + \frac{1}{4i}tr(S[S_X,S_Y]) = 0 \eqno(56b)
$$
where $X=x/2, \quad Y = y/\alpha , \quad w = - \alpha^{-1}u_{Y}$.

(ii) {\it The Ishimori equation.} Now let  $ a=b=-\frac{1}{2} $. Then equation (4) reduces to the known
Ishimori equation [15]
$$
iS_t+\frac{1}{2}[S,(S_{xx}+\alpha^{2}S_{yy})]+
iu_{y}S_x+iu_{x}S_y = 0 \eqno(57a)
$$
$$
\alpha^{2}u_{yy} - u_{xx}= \frac{\alpha^{2}}{2i}tr(S[S_x,S_y]). \eqno(57b)
$$

(iii) {\it The Myrzakulov XXXIV (M-XXXIV) equation.}  This equation has
the form
$$
iS_t+\frac{1}{2}[S,S_{YY}]+iwS_Y = 0 \eqno(58a)
$$
$$
w_{t} + w_{Y} + \frac{1}{4}\{tr(S^{2}_{Y})\}_{Y} = 0. \eqno(58b)
$$

The  M-XXXIV  equation (19)
was proposed in [16] to describe nonlinear dynamics of
compressible magnets. It is integrable and has  the different soliton
solutions [24].

(iv) {\it The Myrzakulov XVIII (M-XVIII) equation.}
Now we  consider the reduction: $a=-\frac{1}{2}$. Then the equation (55)
reduces to the M-XVIII equation [16]
$$
iS_t+\frac{1}{2}[S, S_{xx} + 2\alpha(2b+1)S_{xy}+\alpha^{2}S_{yy}]+
A^{\prime}_{2}S_x+A^{\prime}_{1}S_y = 0 \eqno(59a)
$$
$$
\alpha^{2}u_{yy} - u_{xx}=\frac{\alpha^{2}}{2i}tr(S[S_{x},S_{y}]) \eqno(59b)
$$
where $A^{\prime}_{j} = A_{j}$ as $a=-\frac{1}{2}$.

(v) {\it The Myrzakulov XIX (M-XIX) equation.}
Let us consider the case: $a = b$. Then we obtain the M-XIX equation [16]
$$
iS_t + \frac{1}{2} [S,  \alpha^{2} S_{yy} - 4a(a+1) S_{xx}]
+  A_{2}^{\prime \prime } S_x + A_{1}^{\prime \prime }  S_y = 0  \eqno(60a)
$$
$$
M_{2} u = \frac{\alpha^{2}}{2i} tr( S [S_{x}, S_{y}]) \eqno(60b)
$$
where $A_{j}^{\prime \prime} = A_{j}$ as $ a = b$.

(vi) {\it The Myrzakulov XX (M-XX) equation.} This equation has the form [16]
$$
iS_t +  \frac{1}{2}[S,(b+1) S_{\xi \xi} -bS_{\eta \eta}] +
ibw_{\eta} S_{\eta} + i(b+1)w_{\xi}S_{\xi} = 0 \eqno(61a)
$$
$$
w_{\xi \eta} =  \frac{1}{4i}tr(S[S_{\xi},S_{\eta}]) \eqno(61b)
$$
and so on [16]. The gauge equivalent counterparts of equations (56),
(57), (58) and (61) are the equations (3), (2), (4) and (11), respectively.
Note that from (61) as $b=0$, we get the M-VIII equation in the following
form [16]
$$
iS_t +  \frac{1}{2}[S, S_{\xi \xi}] +  ibw_{\xi}S_{\xi} = 0 \eqno(62a)
$$
$$
w_{\xi \eta} =  \frac{1}{4i}tr(S[S_{\xi},S_{\eta}]) \eqno(62b)
$$
the gauge equivalent of which is the equation (12). If we put $\eta = t$,
then equations (61) and (12) take the forms
$$
iS_t + \frac{1}{2}[S, S_{\xi\xi}]+iwS_{\xi} = 0   \eqno (63a)
$$
$$
w_{t} +  \frac{1}{4}(trS^{2}_{\xi})_{\xi} = 0  \eqno (63b)
$$
and
$$
iq_{t}+q_{\xi \xi}+vq=0 \eqno(64a)
$$
$$
v_{t} + 2r^{2}(\bar q q)_{\xi}=0. \eqno(64b)
$$
Equation (63) is the equivalent form of the M-XXXIV  equations. At the
same time, its gauge equivalent  (64) is the Ma equation [20],
which is also the
equivalent form of the YOE (4).

Note that these spin systems admit the different types solutions (see,
e.g. [21-25]).
\section{ Conclusion}

We have investigated the integrability aspects of the (2+1)-dimensional
ZE by the singularity structure analysis and shown that it admits the
Painleve property. We have also derived its bilinear form directly from
the Painleve analysis. We have then generated the simplest 1-SS using
the Hirota method. We have constructed and the (1,1)
dromion solution.   Finally, in last section we have presented the associated
integrable spin systems, which are gauge equivalent counterparts of the ZE and
its reductions. Here, we would like note that between these spin systems and
the NLSE-type equations can take place the so-called Lakshmanan equivalence.
This problem we will consider, in detail, in other places (see, e.g.,
[26-29]).

\end{document}